\documentclass[12pt]{article}
\pdfoutput=1
\usepackage{graphicx}
\begin{document}

\pagestyle{empty}

\noindent
{\bf Period Analysis of All-Sky Automated Survey for Supernovae (ASAS-SN) Data on Pulsating Red Giants}

\bigskip

\noindent
{\bf John R. Percy and Lucas Fenaux\\Department of Astronomy and Astrophysics, and\\Dunlap Institute of Astronomy and Astrophysics\\University of Toronto\\Toronto ON\\Canada M5S 3H4}

\bigskip

{\bf Abstract}  The All-Sky Automated Survey for Supernovae (ASAS-SN) has
recently used over 2000 days of data to identify more than 50,000 variable stars, automatically classify these, 
determine periods and amplitudes for those that are periodic -- part of a
remarkable project to classify 412,000 known variable stars, and determine
their basic properties.  This information about the newly-discovered variables,
along with the
photometric data is freely available on-line.  In this paper, we analyze ASAS-SN V
data on two small random samples of pulsating red giants (PRGs) in detail, and compare our
results with those found by ASAS-SN.  For the majority of a sample of 29 mostly
semi-regular (SR) PRGs, the ASAS-SN results are incorrect or incomplete:
either the ASAS-SN periods are 2, 3 or 4 times the actual period, or the
ASAS-SN period is a ``long secondary period" with a shorter pulsation period present, or the star is multi-periodic
or otherwise complex, or  the star's data are contaminated by
instrumental effects.  For almost all of a sample of 20 of the {\it longest-period}
Mira stars (period 640 days or more), the ASAS-SN period is actually 2 or more times the actual period.  Our results are not surprising, given the very complex
behaviour of PRGs.

\medskip

\noindent
AAVSO keywords = Photometry, CCD; pulsating variables; giants, red; period analysis; amplitude analysis

\medskip

\noindent
ADS keywords = stars; stars: late-type; techniques: photometric; methods: statistical; stars: variable; stars: oscillations

\medskip

\noindent
{\bf 1. Introduction}

\smallskip

Red giant stars are unstable to pulsation.  In the {\it General Catalogue
of Variable Stars} (GCVS; Samus {\it et al.} 2017), pulsating red giants
(PRGs) are classified according to their light curves.  Mira (M) stars
have reasonably regular light curves, with visual ranges greater than 2.5
magnitudes.  Semi-regular (SR) stars are classified as SRa if there is
appreciable periodicity, and SRb if there is little periodicity.
Irregular (L) stars have very little or no periodicity.

Mira stars have periods which ``wander" by a few percent; this wandering
can be described and modelled as random, cycle-to-cycle fluctuations (Eddington and
Plakidis 1929).  Their maximum magnitudes vary from cycle to cycle, as
observers of Mira itself know.  The variability of SR stars is even
more complicated.  Some stars are multiperiodic; both the fundamental
and first overtone modes are excited (e.g. Kiss {\it et al.} 1999).  About a third show long secondary
periods (LSPs), 5-10 times the pulsation period (Wood 2000); their cause
is unknown.  The amplitudes of PRGs vary by up to a factor of 10 on time
scales of 20-30 pulsation periods (Percy and Abachi 2013).  There are also
very slow variations in mean magnitude (Percy and Qiu 2018).  In a very few stars, thermal pulses
cause large, secular changes in period, amplitude, and mean magnitude
(Templeton {\it et al.} (2005) and references therein).

Our previous studies of PRGs have used long-term visual and sometimes
photoelectric observations from the American Association of Variable
Star Observers International Database (AID; Kafka 2019).  Now, an
important and very useful new source of data is available: the All-Sky Automated
Survey for Supernovae (ASAS-SN).

ASAS-SN uses a network of up to 24 telescopes around the world to survey
the entire visible sky every night down to about 18th magnitude
(Shappee {\it et al.} (2014), Jayasinghe {\it et al.} (2018)).  It has
been doing so for over 2000 days (since about JD 2456500).  ASAS-SN has identified over 50,000
variable stars, classified these using machine learning, determined periods and amplitudes for
those that are periodic, ane made this information, and the data on-line
(asas-sn.osu.edu/variables).  It has also used machine learning to uniformly
classify 412,000 known variables (Jayasinghe {\it et al.} 2018).

The purpose of the present project was to look at a small, random
sample of the PRG data in more
detail, and investigate the reliability of the ASAS-SN classifications and
periods and amplitudes of PRGs.  Jayasinghe {\it et al.} (2018) comment
only briefly on the ASAS-SN classification of these very complex variables.

Vogt {\it et al.} (2016) have recently carried out a related study: analysis of
2875 Mira stars observed in the original ASAS project, which extended
from 2000 to 2009.  They used a semi-automatic method based on the observed times of
maximum light.  They found that, whereas their periods agreed with those
in the VSX Catalogue (Watson {\it et al.} 2014) in more than 95 percent of the stars, their periods
agreed with those obtained by Richards {\it et al.} (2012), who used an automatic
machine-learning method, in only 76 percent of the stars.  Most often,
the latter periods differed from the Vogt {\it et al.} (2016) periods by a ratio
of small whole numbers.

\medskip

\noindent
{\bf 2. Data and Analysis}

\smallskip

For our initial project, we analyzed a sample of 22 stars classified
as SR, five as M, and three as L.  They were randomly chosen around a random
position on the sky.  The SR and L stars were chosen to have ASAS-SN amplitudes
of at least 0.5 magnitude (with one accidental exception), so that the results would not be unduly affected
by noise, and would therefore be meaningful.
A study of smaller-amplitude ASAS-SN stars would be the subject of a
future project. The datasets were approximately 2000 days in length,
significantly shorter than visual datasets from the AID.  For each of the
SR and M stars, ASAS-SN provides a period and amplitude, and a light
curve and phase curve, and a quantity T which is a statistical measure
of the confidence of the period; lower values indicate higher confidence. 

The data were downloaded, and analyzed using the
AAVSO VSTAR time-series package (Benn 2013) which includes a Fourier
analysis routine.  For one star, we also used a new self-correlation (e.g. Percy {\it et al.} 2003) program
written by author LF.

The results were interesting, so we carried out a subsidiary project,
to analyze a sample of 20 Mira stars with the longest ASAS-SN periods -- longer
than 639 days.  Miras with such long periods would be 
especially interesting and important, astrophysically.

\medskip

\noindent
{\bf 3 Results}

\smallskip

Of the 29 stars in our initial project, 7 were acceptably analyzed (e.g. Figure 1).  For 9 stars,
the ASAS-SN period was exactly 2, 3, or 4 times the actual period; the
phase curve had not one, but 2, 3, or 4 cycles in it.  While this is a
mathematical possibility, it is unphysical for a radial pulsator such as a 
PRG.  Figure 2 shows an example with 3 cycles per unit phase.

For 5 stars, the light curve also showed periodic variability on a time scale
5-10 times shorter than the ASAS-SN period.  The latter was clearly a ``long
secondary period", whereas the shorter period was the pulsation period.
Figures 3-5 show an example in which the LSP is actually half the ASAS-SN period of 1022 days.  The
shorter pulsation period of 55$\pm$2 days is clearly visible.

For a very few stars, the light curve included some highly-discordant data, 
and it appeared that ASAS-SN had used all the data for analysis.  Figure 6 
shows an example.  These discordant photometric points are probably due
to astrometry problems, and their effect on the image-subtraction process (Kochanek and Jaysinghe, private communication, April 2, 2019).  The non-discordant data show periods of 423.9 and 66.9 days, with V 
amplitudes
of 0.09 and 0.07, respectively, rather than the (artificial) ASAS-SN amplitude of 2.5.  The longer period is probably a long secondary period.

For a few other stars, the variability appears to be either bimodal or more complex.
Figure 7 shows an example in which there may be periods of 469 days (the
ASAS-SN period) and about half that value -- typical of PRGs which are
pulsating in the fundamental and first overtone modes.  Bimodal pulsators can be useful for
determining the physical properties of the stars.

Figure 8 shows the light curve of a star which ASAS-SN classifies as irregular
(L type) but which clearly shows some periodicity; we obtain a best period of
121 days.

Figure 9 shows a star with a very unusual light curve.  There are two
maxima which take the form of slow ``eruptions".  They may, however, be
maxima of a faint Mira star which has a 15.3-magnitude non-variable
companion.
 
Table 1 lists the results of the initial project.  It  gives: the ASAS-SN
name of the star, minus ASAS-SN-V J; the period PA in days given by ASAS-SN;
the pulsation period PP in days obtained by us; the V amplitude $\Delta$V; the mean
V magnitude $<$V$>$; the (J-K) color; the ASAS-SN classification; and the following notes: x2, x3,
x4: the ASAS-SN period is 2, 3, or 4 times the correct pulsation period;
lsp: the ASAS-SN period is a long secondary period, and a shorter pulsation
period can be seen in the light curve; tpp: the star shows evidence of two
pulsation periods, differing by a factor of approximately two; dd: the analysis
is affected by contamination by discordant data (see above); spp: the
pulsational phase curve is more sawtooth than sinusoidal; OK: the ASAS-SN analysis is
correct; *: see ``Notes on Individual Stars"
below.  This Table, and Figures 1-9 show the remarkable diversity of
results which occur in a sample of less than 30 stars.

In the 20 long-period Mira stars in the subsidiary project, the ASAS-SN
period was in almost every case exactly 2, 3, 4, or 5 times the
actual period.
 
Table 2 lists 20 Mira stars with mean V magnitudes between 12 and 14, and
with the longest periods.  The magnitude range was chosen because it is optimal for ASAS-SN data.  They are listed in order of decreasing ASAS-SN
period.  The columns list: the name of the star, minus ASAS-SN-V J; the
period PA in days given by ASAS-SN; the mean V magnitude $<$V$>$; the
ASAS-SN amplitude $\Delta$VA; and the following
notes: x2, x3, x4, x5: the ASAS-SN period is 2, 3, 4, or 5
times the correct pulsation period; dd: the analysis seems to
have been complicated by discordant data (see above); OK: the
ASAS-SN analysis is correct; *: see
``Notes on Individual Stars", below.

\medskip

\noindent
{\bf 3.1 Notes on Individual Stars in Table 1}

\smallskip

Figures 1-9 and their captions provide both light/phase curves and notes about
seven illustrative stars in the sample.

\smallskip

{\it ASAS-SN-V J054606.99-694202.8:}  The light curve is unusual; it is
non-sinusoidal, and there are two maxima in the 544-day cycle.  It is not
clear whether the behavior is periodic.

\smallskip

{\it ASAS-SN-V J053035.52-685923.2:}  The light curve shows a slow decline, with some cyclic variations superimposed; their time scale is about 200 days.  The
slow decline could be part of a long secondary period.

\smallskip

{\it ASAS-SN-V J052011.96-694029.4:}  The light curve (Figure 9) seems to show outbursts, but they may be maxima of a faint Mira star which has a non-variable
companion.

\smallskip

{\it ASAS-SN-V J054110.62-693804.1:}  The star has a double-humped maximum.

\smallskip

{\it ASAS-SN-V J045337.64-691811.2:}  Unlike the other stars in the sample,
this star had a very small amplitude, but it was possible to show that the
actual period is 1/5 of the ASAS-SN period.

\smallskip

{\it ASAS-SN-V J 185653.55-392537.4:}  The pulsation amplitude is slowly
decreasing during the time of observation.

\smallskip

{\it ASAS-SN-V J042630.05+255344.6:}  One-half the ASAS-SN period is a long secondary period.  A shorter pulsation period is also present.

\medskip

\noindent
{\bf 3.2 Notes on Individual Stars in Table 2}

\smallskip

{\it ASAS-SN-V J171247.59+265024.8:}  There are a few points between
JD 2457850-2457896 which are four magnitudes fainter than the rest, which
are almost constant; these are presumably due to instrumental effects, as
discussed above.
For the rest of the points,
the highest peak has an amplitude of only 0.017 mag.

\smallskip

{\it ASAS-SN-V J195424.95-114932.2:}  There are discordant points.
For the rest, 
the highest peak is at 423.9 days (half the ASAS-SN period) with
an amplitude of 0.09 mag; the second-highest is at 66.9 days, with an
amplitude of 0.07 mag.  These may be an LSP and a pulsation period.  The
ASAS-SN amplitude is given as 2.5 magnitudes.

\smallskip

{\it ASAS-SN-V J181958.07-395457.8:}: there are a few discordant points.

\smallskip

{\it ASAS-SN-V J182346.68-363942.1:}  There are discordant points.  For the rest,
periods of 158$\pm$8 days and 83$\pm$4 days are present, with small amplitudes (Figure 10).  They may possibly be the fundamental and first overtone pulsation periods.

\smallskip

{\it ASAS-SN-V J144304.69-753418.9:} The light curve is unusual; there are
variations on a time scale of about 100 days, superimposed on irregular long-term variations (Figure 11).  The ASAS-SN period of 641.8 days is unlikely.
\medskip
 
\medskip

\noindent
{\bf 4. Discussion}

\smallskip

The ASAS-SN data begin about JD 2456500 so, as of the time of carrying out 
this project, there is only about 2000 days of data.  This is adequate for studying
many aspects of PRG variability, but not the very long-term variations in period,
amplitude, and mean magnitude.  Only the visual data can presently do
that.  Nevertheless, these new data provide a remarkable
resource for studying these and other variable stars.

It is interesting to note that, when Vogt {\it et al.} (2016) compared their
results with those of Richards {\it et al.}'s (2012) results which were obtained using a
machine-learning approach, the discrepancy was most often by a ratio of
small whole numbers, such as 2 or 1/2.  We find a similar result.

Pulsating red giants are certainly a challenge for automated analysis
and classification.  Jayasinghe (private communication, April 10, 2019) has
been refining the analysis and classification procedure, and has provided
a list of updated periods for the stars in Table 1.  About two-thirds now
agree with our values, so a detailed inspection and analysis of the light
curve is still recommended for individual SR and Mira stars.

It is not surprising that a few of these stars, especially the longer-period
stars, have sawtooth phase curves.  The same was found by Percy and Qiu (2018),
from AAVSO visual data.

Most of this project was carried out by undergraduate math major LF.  It
illustrates the great educational potential of the ASAS-SN data, with its
immense quantity, quality, and variety. We can envision a large number
and variety of projects which could be carried out by students, using
the ASAS-SN data.  The AAVSO VSTAR time-series
analysis package is well-suited for use with these and other data.

\medskip

\noindent
{\bf 5. Conclusions}

\smallskip

We have analyzed ASAS-SN observations of pulsating red giants (mostly semi-regular
and Mira stars) and compared our results with the periods, amplitudes, and
classifications given by ASAS-SN.  For many stars, the actual periods are
a small integral fraction of the ASAS-SN period, because the ASAS-SN phase
curve contains two or more cycles of variability, rather than one.  In other cases, the
ASAS-SN period is a long secondary period, and the shorter pulsation period
is visible in the light curve.  In a few stars, the ASAS-SN analysis is
complicated by the presence of data which are discordant, due to instrumental
problems.  In a few others, the star is
bimodal or otherwise complex.  The few irregular (type L) stars that we
analyzed were probably semi-regular (type SR).

Given the complexity of pulsating red giants, it is not surprising that the
ASAS-SN automatic analysis procedure produced incorrect or incomplete
results.  Perhaps the procedure can be trained to ``solve" these very
complex stars!  Indeed, the ASAS-SN variable star data and website have been
significantly updated and improved in the weeks since we completed this
project (in February 2019), and some of the problems with the PRG analysis
and classification have been fixed.  

The ASAS-SN data on PRGs can be exceptionally useful for
analyzing these stars, and is invaluable for both scientific and educational
purposes.  But the data for individual PRGs in the ASAS-SN catalogue should still be confirmed
by careful inspection of the light curve, and by detailed analysis if
necessary.

\medskip

\noindent
{\bf Acknowledgements}

\smallskip

This paper made use of ASAS-SN photometric data.
We thank the ASAS-SN project for their remarkable contribution to
stellar astronomy, and for making the data freely available on-line.  Thanks
also to Chris Kochanek and especially Tharindu Jayasinghe for helpful comments.
We acknowledge and thank the University
of Toronto Work-Study Program for financial support.
The Dunlap
Institute is funded through an endowment established by the David Dunlap
Family and the University of Toronto.

\bigskip

\noindent
{\bf References}

\smallskip

\noindent
Benn, D. 2013, VSTAR data analysis software (http://www.aavso.org/node/803)

\smallskip

\noindent
Eddington, A.S. and Plakidis, S. 1929, {\it Mon. Not. Roy. Astron. Soc.},
{\bf 90}, 65.

\smallskip

\noindent
Jayasinghe, T. {\it et al.} 2018, {\it ArXiv:} 1809.07329

\smallskip

\noindent
Kafka, S. 2019, variable star observations from the AAVSO International Database

(https://www.aavso.org/aavso-international-database)

\smallskip

\noindent
Kiss, L.L., Szatmary, K., Cadmus, R.R., Jr., and Mattei, J.A. 1999,
{\it Astron. Astrophys.}, {\bf 346}, 542.

\smallskip

\noindent
Percy, J.R., Hosick, J., and Leigh, N. 2003, {\it Publ. Astron. Soc. Pacific}, {\bf 115}, 59.

\noindent
Percy, J.R. and Abachi, R. 2013, {\it J. Amer. Assoc. Var. Star Obs.},
{\bf 41}, 93.

\smallskip

\noindent
Percy, J.R. and Qiu, A. 2018, arXiv: 1805.11027

\smallskip

\noindent
Richards, J.W., {\it et al.} 2012, {\it Astrophys. J. Suppl.}, {\bf 203}, 32.

\smallskip

\noindent
Samus, N.N. {\it et al.} 2017, {\it General Catalogue of Variable Stars},
Sternberg Astronomical Institute, Moscow (GCVS database: http://www.sai.msu.ru/gcvs/gcvs/index.htm)

\smallskip

\noindent
Shappee, B.J. {\it et al.} 2014, {\it Astrophys. J.}, {\bf 514}, 932.

\smallskip

\noindent
Templeton, M.R., Mattei, J.A. and Willson, L.A. 2005, {\it Astron. J.},
{\bf 130}, 776.

\smallskip

\noindent
Vogt, N. {\it et al.} 2016, {\it Astrophys. J. Suppl.}, {\bf 227:6}, 1. 

\smallskip

\noindent
Watson, C., Henden, A.A., and Price, C.A. 2014, www.aavso.org/vsx

\smallskip

\noindent
Wood, P.R. 2000, {\it Publ. Astron. Soc. Australia}, {\bf 17}, 18.

\medskip

\smallskip

\begin{table}
\begin{center}
\caption{Table 1.  Analysis of ASAS-SN Observations of 29 Pulsating Red Giants}
\begin{tabular}{rrrrrrrl}
\hline
Name - ASAS-SN-V & Type & PA(d) & P(d) & $\Delta$V & $<$V$>$ & J-K & Notes (see text) \\
\hline
J053227.48-691652.8 & SR & 469 & 227 & 1.2 & 12.7 & 1.112 & x2?,  Fig. 7 \\
J055444.75-694714.7 & SR & 544 & 264 & 1.5 & 12.86 & 0.951 & x2, lsp \\
J061214.08-694558.6 & SR & 643 & 318 & 1.5 & 16.82 & 1.725 & x2 \\
J054102.00-704309.9 & SR & 702 & 702 & 1.2 & 15.78 & 1.357 & OK, spp \\
J191920.70-195042.1 & SR & 82 & 81 & 0.8 & 13.43 & 1.201 & OK, tpp? \\
J054747.21-602210.3 & SR & 418 & 139 & 2.5 & 13.27 & 0.944 & x3, Fig. 2 \\
J191639.10-215848.8 & SR & 25 & 38 & 0.3 & 11.61 & 1.245 & \\
J205350.26-593921.1 & SR & 168 & 168 & 1.2 & 11.81 & 1.209 & tpp \\
J054606.99-694202.8 & SR & 544 & 538 & 1.5 & 15.92 & 1.381 & OK, spp, * \\
J191715.66-200034.1 & SR & 59 & 59 & 0.8 & 12.75 & 1.191 & OK \\
J051623.43-690014.3 & SR & 466 & 233 & 0.7 & 14.95 & 1.287 & x2 \\
J053035.52-685923.2 & SR & 643 & 295 & 0.9 & 13.13 & 1.194 & lsp?, tpp, * \\
J052011.96-694029.4 & SR & 662 & 662 & 1.1 & 15.1 & 1.494 & OK, spp, Fig. 9, * \\
J052337.99-694445.8 & SR & 636 & 400 & 1.2 & 16.38 & 1.228 & \\
J054036.77-692620.6 & SR & 505 & 458 & 1.0 & 13.39 & 1.172 & tpp \\
J054110.62-693804.1 & SR & 702 & 694 & 1.4 & 12.65 & 1.158 & OK, * \\
J045337.64-691811.2 & SR & 430 & 80 & 0.2 & 13.3 & 1.115 & x5, * \\
J045412.77-701708.6 & SR & 437 & 204 & 0.5 & 13.58 & 1.181 & tpp \\
J050354.98-721652.3 & SR & 138 & 138? & 1.5 & 16.21 & 0.898 & OK, Fig. 1 \\
J171247.59+265024.8 & M & 889 & -- & 0.0 & 13.4 & 0.785 & \\
J195424.95-114932.2 & M/SR & 848 & 424 & 0.2 & 13.5 & 1.269 & lsp, Fig. 6 \\
J175514.90+184006.9 & M & 815 & 204 & 2.7 & 13.68 & 1.186 & x4 \\
J182825.60+171943.2 & M & 728 & 243 & 2.35 & 13.23 & 1.538 & x3 \\
J185653.55-392537.4 & M & 645 & 215 & 2.30 & 13.76 & 1.272 & x3, * \\
J020359.53+141132.4 & L/SR & irr & 389 & 1.25 & 12.01 & 1.148 & SR, lsp? \\
J181616.35-281634.1 & L/SR & irr & 121 & 1.38 & 13.69 & 1.149 & SR, Fig. 8 \\
J194755.85-611127.5 & L/SR & irr & 400 & 1.14 & 13.07 & 1.151 & SR \\
J042630.05+255344.6 & SR & 1022 & 30 & 0.95 & 13.66 & 1.759 & x2, lsp, Fig. 3-5, * \\
J082819.18-143319.3 & SR & 1020 & 78/128 & 0.63 & 11.66 & 1.251 & tpp \\ 
\hline
\end{tabular}
\end{center}
\end{table}

\begin{table}
\begin{center}
\caption{Table 2.  Analysis of ASAS-SN Observations of 20 Long-Period Mira Stars}
\begin{tabular}{rrlll}
\hline
Name - ASASSN-V & PA(d) & $<$V$>$ & $\Delta$VA & Notes (see text) \\
\hline
J171247.59+265024.8 & 888.8 & 13.5 & 3.05 & * \\
J195424.95-114932.2 & 848.4 & 13.6 & 2.5 & lsp?, * \\
J175514.90+184006.9 & 814.7 & 13.7 & 2.7 & x4 \\
J182825.60+171943.2 & 814.7 & 13.2 & 2.35 & x3 \\
J065708.96+473521.9 & 725.8 & 13.48 & 2.02 & x2, QX Aur \\
J202918.27+125429.1 & 721.0 & 13.31 & 2.22 & x5, XZ Del \\
J190214.90+471259.7 & 716.1 & 13.49 & 2.71 & x2, WZ Lyr \\
J175727.78+243018.0 & 695.1 & 13.31 & 2.56 & x2 \\
J184802.27-293034.0 & 675.7 & 13.27 & 2.36 & dd \\
J184706.22-314645.6 & 675.3 & 13.62 & 4.61 & x3, V962 Sgr \\
J181958.07-395457.8 & 665.9 & 12.75 & 2.78 & * \\
J082915.17+182307.3 & 655.8 & 13.66 & 2.1 & x2 \\
J124209.54-435503.3 & 645.6 & 13.93 & 2.79 & OK, V1132 Cen \\
J182037.28-385833.5 & 645.5 & 13.8 & 2.02 & x4 \\
J182346.68-363942.1 & 645.5 & 13.73 & 2.39 & Fig. 10, * \\
J185653.55-392537.4 & 645.0 & 13.76 & 2.29 & x3, AB CrA, * \\
J144304.69-753418.9 & 641.8 & 12.25 & 2.5 & Fig. 11, * \\
J141547.57-480350.7 & 641.0 & 13.65 & 3.02 & x3 \\
J175730.94-744810.7 & 640.5 & 13.45 & 2.05 & x4 \\
J184614.49-301856.4 & 639.6 & 13.64 & 2.02 & x3, V1935 Sgr \\
\hline
\end{tabular}
\end{center}
\end{table}

\begin{figure}[t]
\vspace{-2cm}
\begin{center}
\includegraphics[height=20cm]{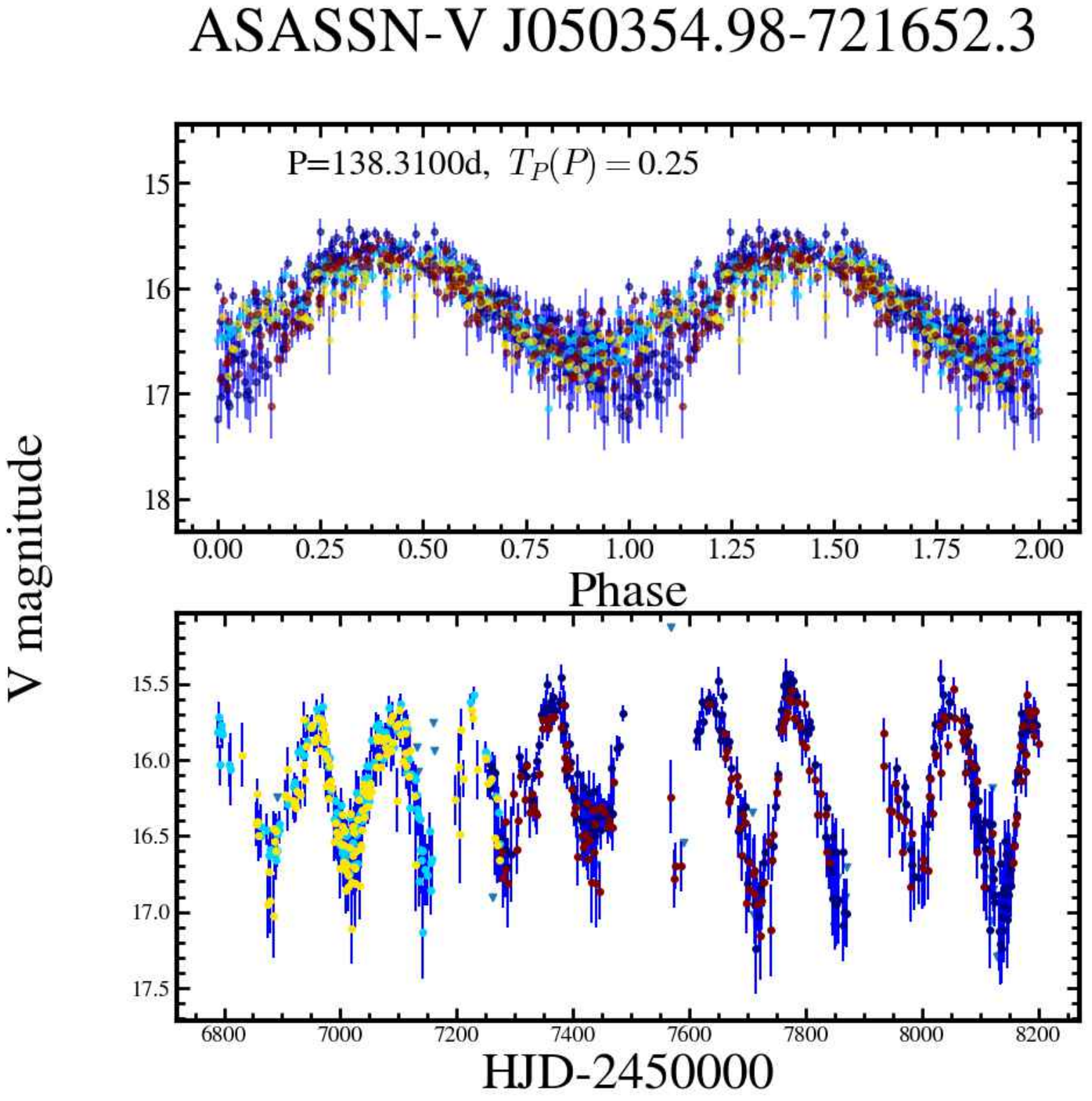}
\end{center}
\vspace{-7cm}
\caption{ASAS-SN-V J050354.98-721652.3: Light curve (bottom), and phase
curve (top) using the ASAS-SN period of 138.3 days.  This period satisfactorily
represents the data.  In this and the following figures, T is a statistical
measure of confidence in the star's period. Source: ASAS-SN website.}
\end{figure}

\begin{figure}[t]
\vspace{-2cm}
\begin{center}
\includegraphics[height=20cm]{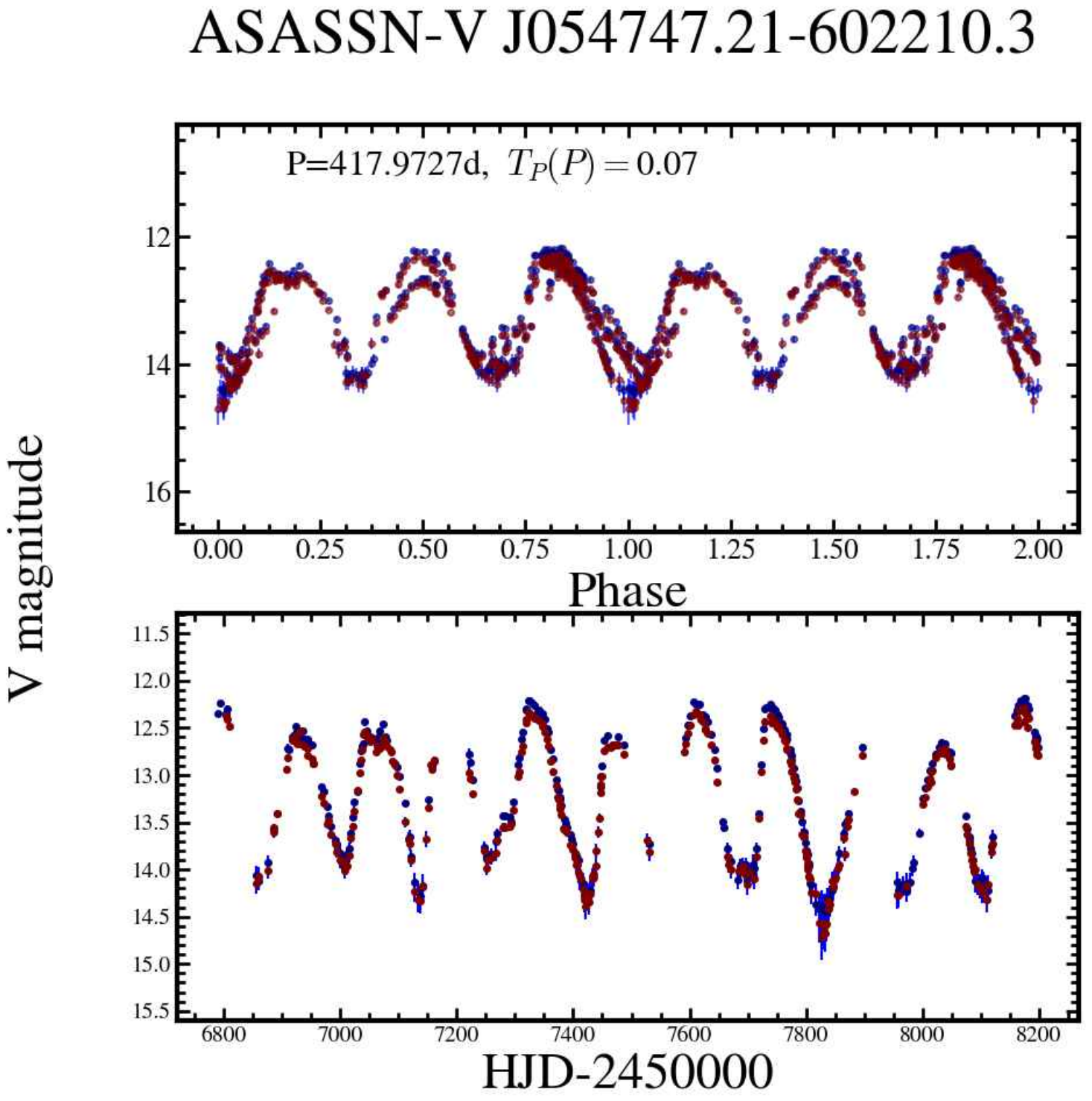}
\end{center}
\vspace{-7cm}
\caption{ASAS-SN-V J054747.21-602210.3: Light curve (bottom), and phase curve
(top) using the ASAS-SN period of 418.0 days. The actual period is one-third
of this; there are three cycles in the phase curve, rather than one.  Source:
ASAS-SN website.}
\end{figure}

\begin{figure}[t]
\vspace{-2cm}
\begin{center}
\includegraphics[height=20cm]{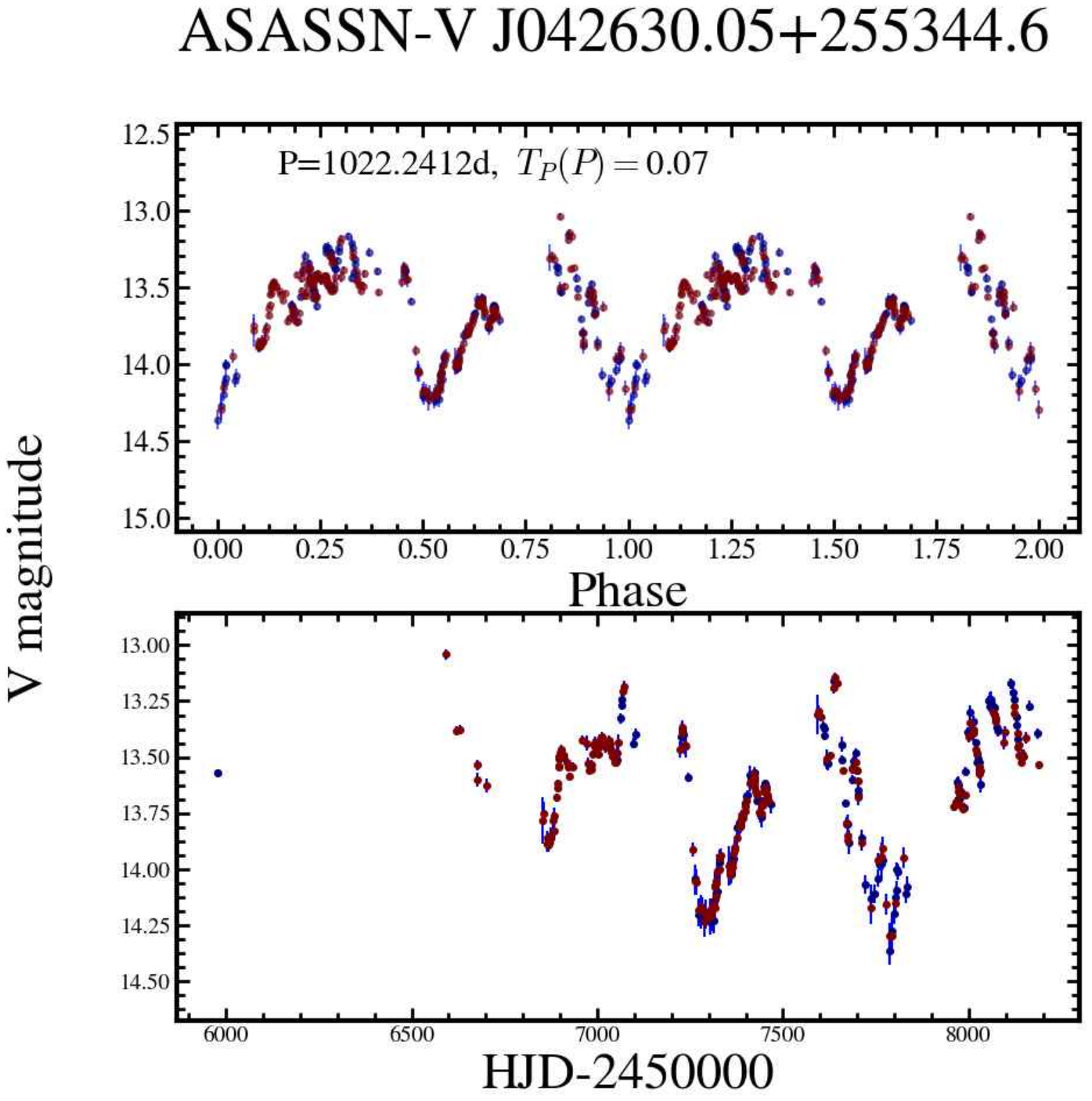}
\end{center}
\vspace{-7cm}
\caption{ASAS-SN-V J042630.05+255344.6: Light curve (bottom), and phase
curve (top) using the ASAS-SN period of 1022.2 days.  There are two (long)
cycles in the phase curve, rather than one, and there are also more rapid
variations with a period of 55$\pm$2 days.  This is presumably the pulsation period,
and the long secondary period is 511.1 days -- half the ASAS-SN period.  Source:
ASAS-SN website.}
\end{figure}

\clearpage

\begin{figure}[t]
\vspace{-0cm}
\begin{center}
\includegraphics[width=15cm]{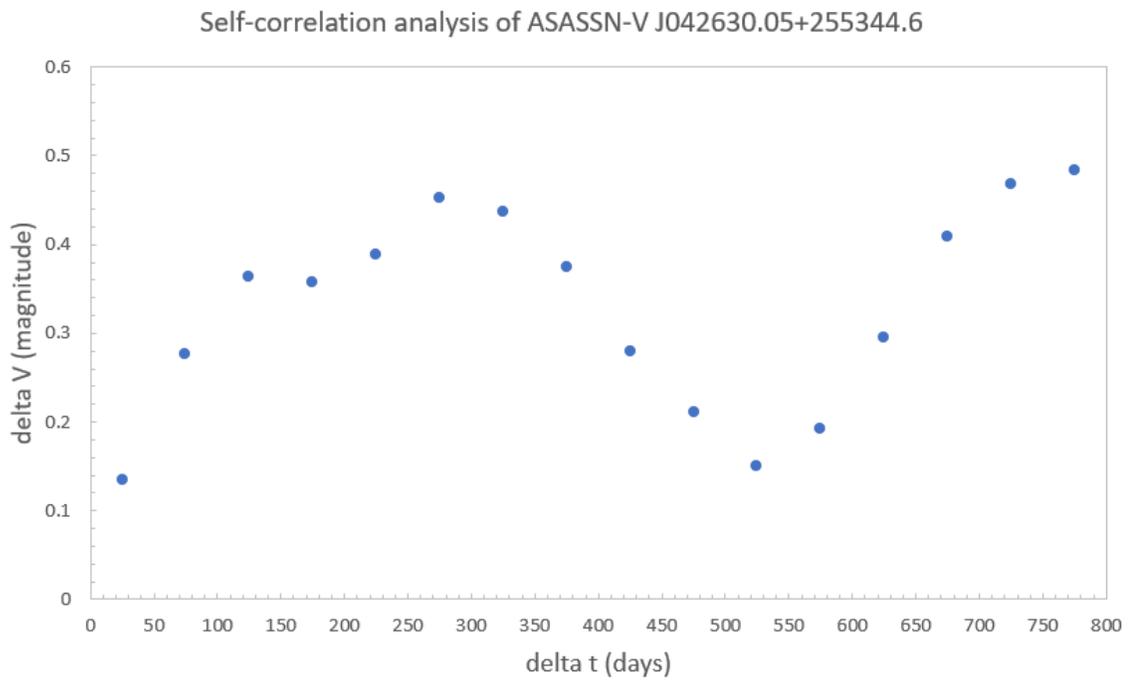}
\end{center}
\vspace{-0cm}
\caption{Self-correlation diagram for ASAS-SN-V J042630.05+255344.6 for
$\Delta$t = 0 to 800 days.  The minimum at about 520 days is the long
secondary period.  The curve does not go to zero at minimum because of the
combined effect of observational error, and the presence of the pulsation
period.}
\end{figure}

\clearpage

\begin{figure}[t]
\vspace{-0cm}
\begin{center}
\includegraphics[width=15cm]{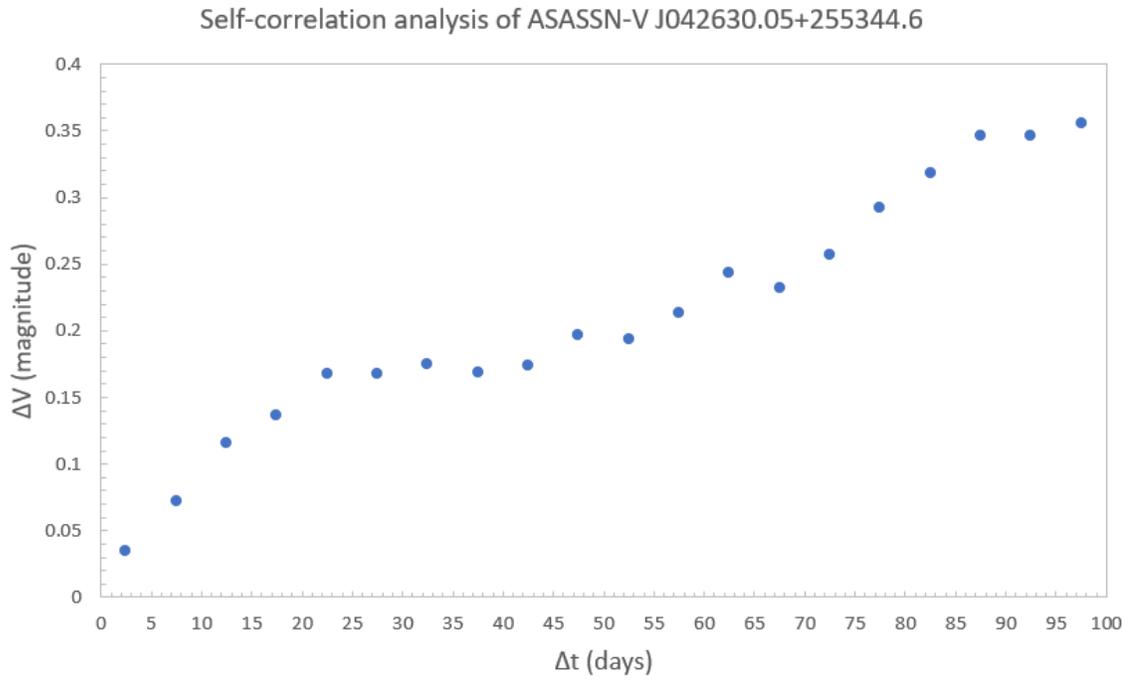}
\end{center}
\vspace{-0cm}
\caption{Self-correlation diagram for ASAS-SN-V J042630.05+255344.6 for
$\Delta$t = 0 to 100 days.  The rising curve is due to the 520-day period;
the shallow minimum (or inflection) at about 50 days is due to the pulsation period.}
\end{figure}

\begin{figure}[t]
\vspace{-2cm}
\begin{center}
\includegraphics[height=20cm]{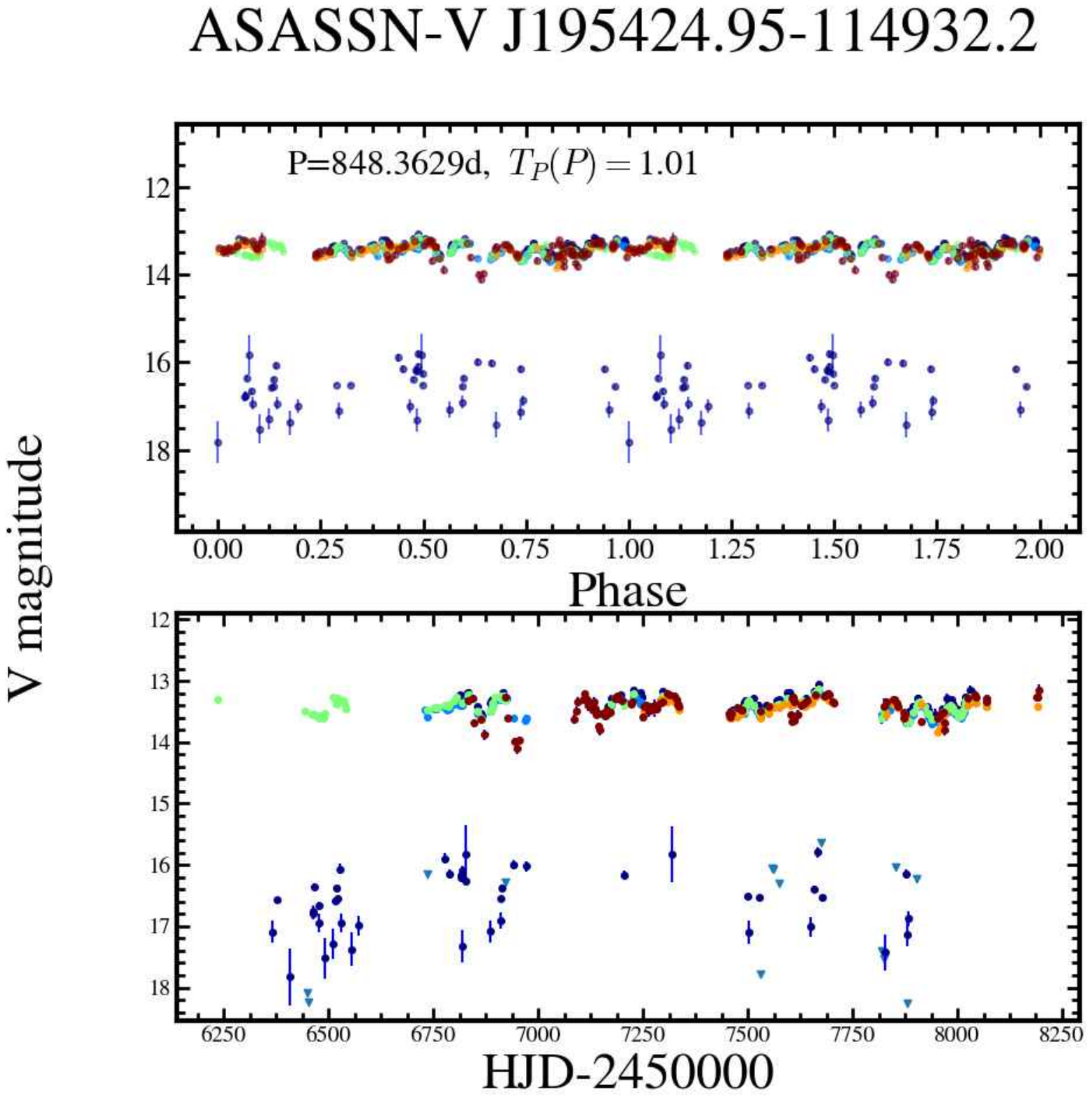}
\end{center}
\vspace{-7cm}
\caption{ASAS-SN-V J195424.95-114932.2; Light curve (bottom), and phase
curve (top) using the ASAS-SN period of 848.4 days.  The ASAS-SN analysis
has been complicated by the fainter discordant points.  Analysis of the brighter
V data gives periods of 423.9 days (V amplitude 0.09) and 66.9 days (V amplitude
0.07).  The former period (half the ASAS-SN period) may be a long secondary
period, and the latter may be a pulsation period.  Source: ASAS-SN website.}
\end{figure}

\begin{figure}[t]
\vspace{-2cm}
\begin{center}
\includegraphics[height=20cm]{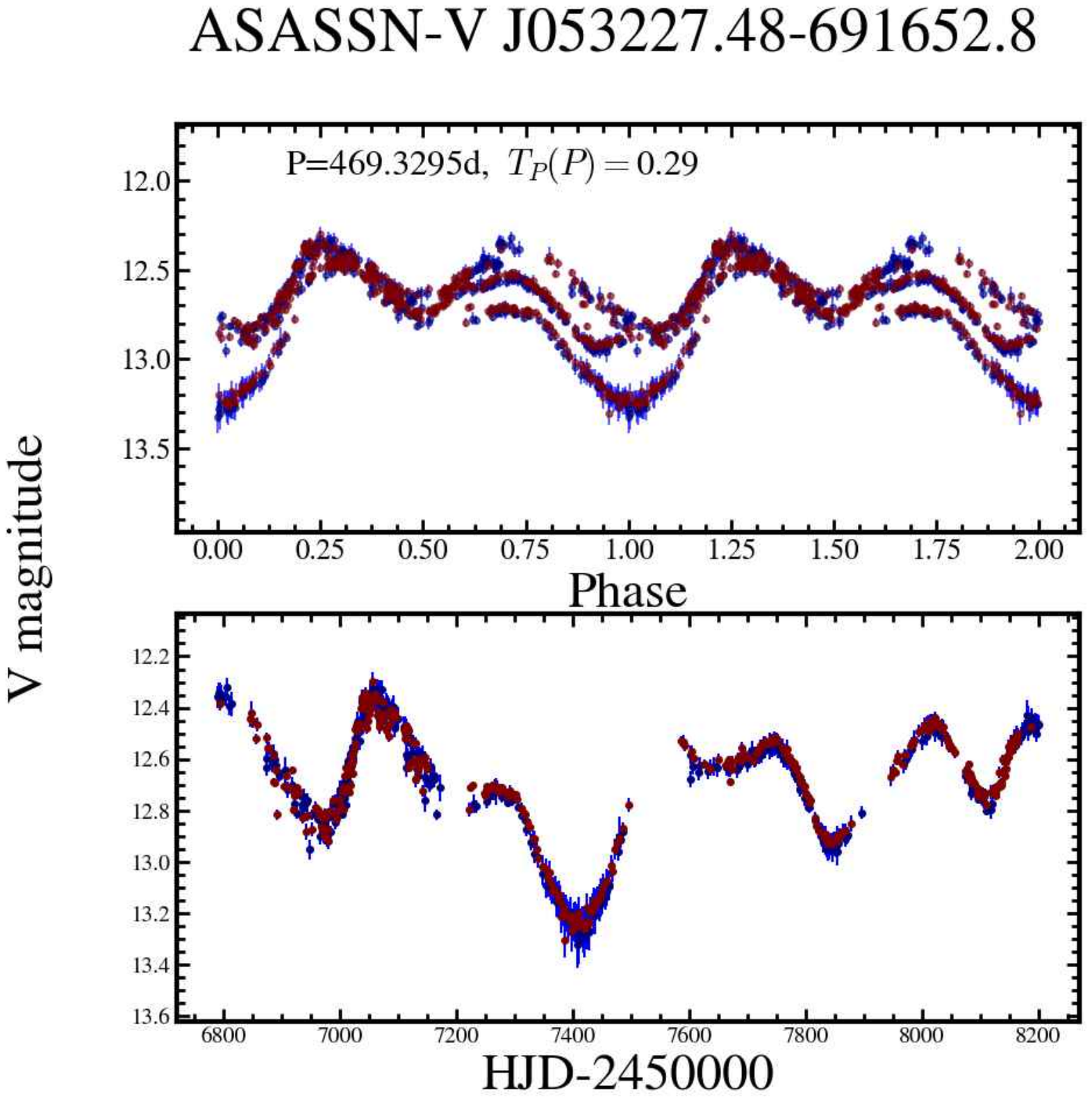}
\end{center}
\vspace{-7cm}
\caption{ASAS-SN-V J053227.48-691652.8: Light curve (bottom), and phase
curve (top) using the ASAS-SN period of 469.3 days.  The star may pulsate
in two modes, with the second period being about half of the first period.
Source: ASAS-SN website.}
\end{figure}

\begin{figure}[t]
\vspace{-2cm}
\begin{center}
\includegraphics[height=20cm]{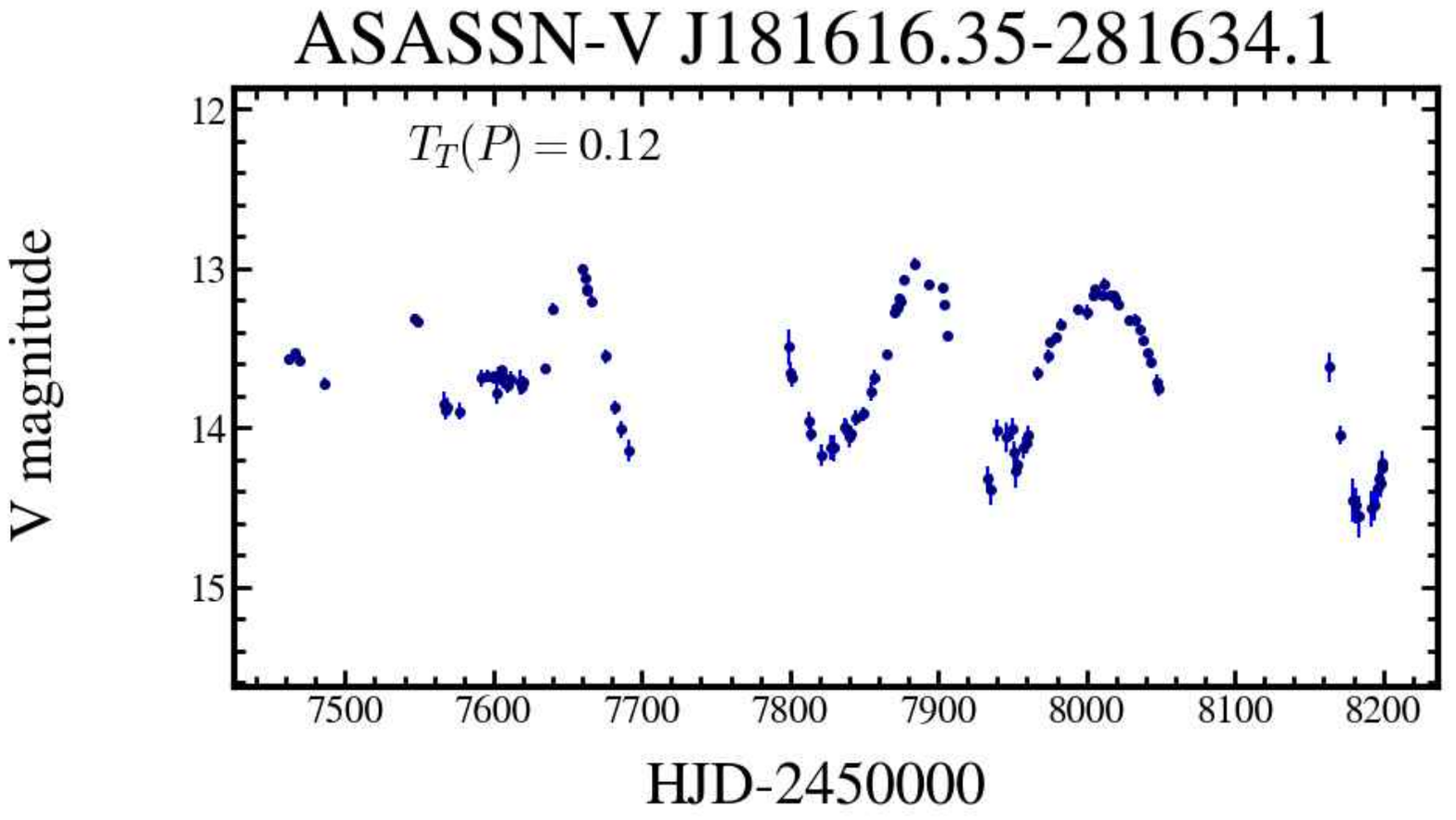}
\end{center}
\vspace{-7cm}
\caption{ASAS-SN-V J181616.35-281634.1: This star is considered irregular
(type L) in the ASAS-SN catalogue, but the above light curve suggests that it
has a period of 121 days, and is therefore an SR star.  Source: ASAS-SN website.}
\end{figure}

\begin{figure}[t]
\vspace{-2cm}
\begin{center}
\includegraphics[height=20cm]{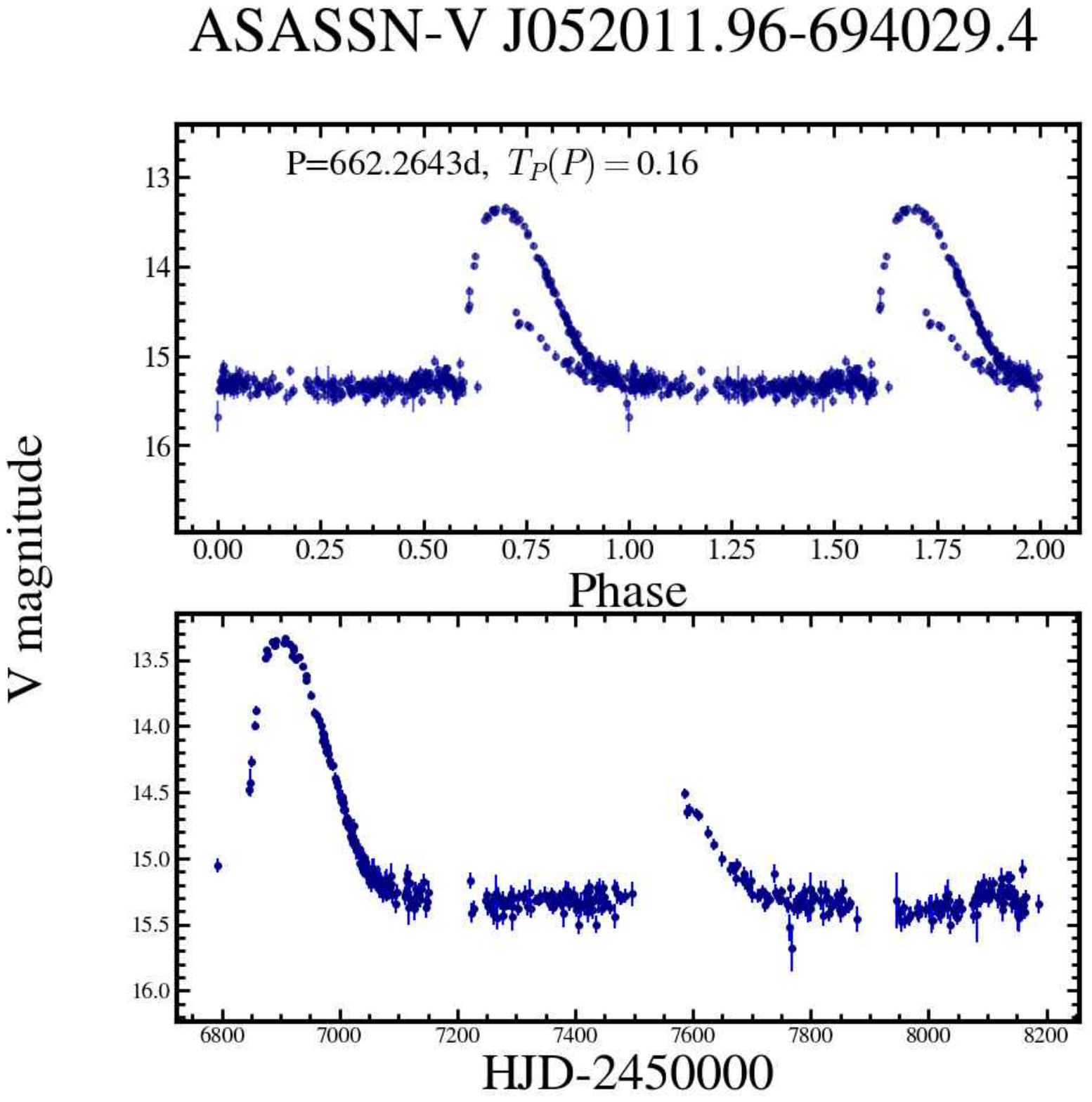}
\end{center}
\vspace{-7cm}
\caption{ASAS-SN-V J052011.96-694029.4: Light curve (bottom), and phase
curve (top) using the ASAS-SN period of 662.3 days.  The light curve shows
two ``eruptions", 662 days apart.  On the other hand, these could be maxima
of a faint Mira star with a constant companion with magnitude 15.3.  Source:
ASAS-SN website.}
\end{figure}

\begin{figure}[t]
% \vspace{-10cm}
\begin{center}
\includegraphics[height=10cm]{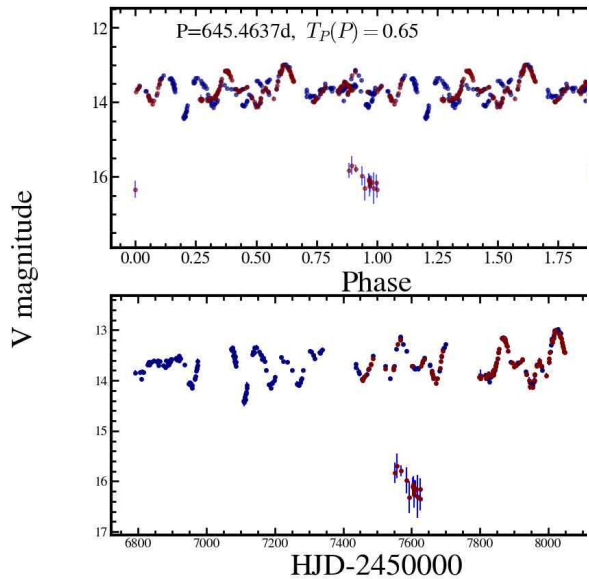}
\end{center}
% \vspace{-7cm}
\caption{ASAS-SN-V J182346.68-363942.1: Light curve (bottom), and phase
curve (top) using the ASAS-SN period of 645.5 days.  The ASAS-SN amplitude
of 2.39 occurs because of the presence of the fainter discordant data.  Our
analysis of the rest of the data gives periods of 153$\pm$8 and 83$\pm$4 days, both
with amplitudes of 0.23.  This may be a bimodal pulsator.  Source: ASAS-SN website.}
\end{figure}

\begin{figure}[t]
% \vspace{-10cm}
\begin{center}
\includegraphics[height=10cm]{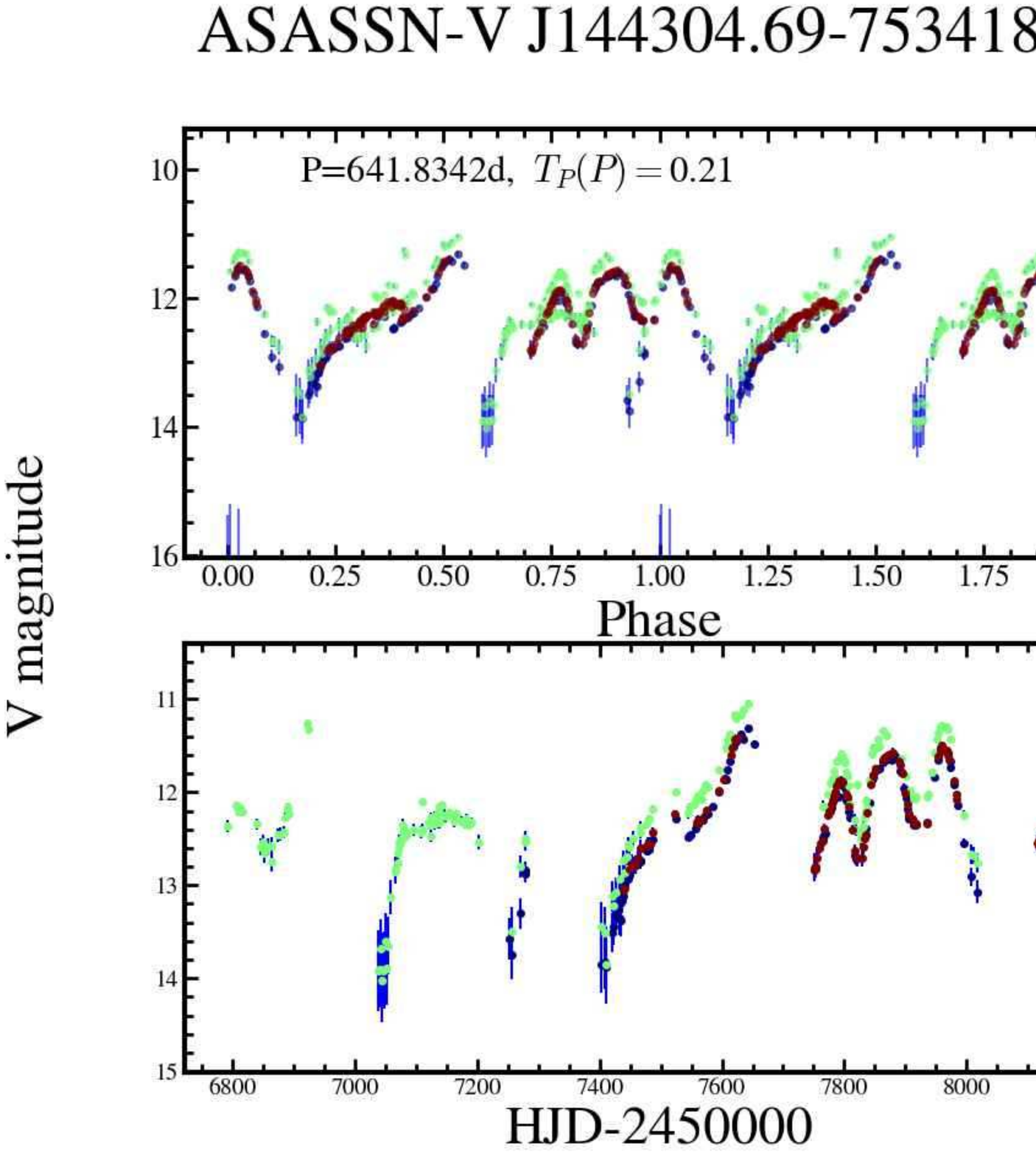}
\end{center}
% \vspace{-7cm}
\caption{ASAS-SN-V J144304.69-753418.9: Light curve (bottom), and phase
curve (top) using the ASAS-SN period of 641.8 days.  The light curve is
highly unusual.  There are variations on time scales from 100-300 days.  Source: ASAS-SN website.}
\end{figure}

\end{document}